\def\md{\mathrm{d}}

\def\bn{{\bf n}}

\def\bs{{\bf s}}

\def\hbn{\hat{\bf n}}

\def\bs1{\textbf{\textsf{1}}}
\def\bsM{\textbf{\textsf{M}}}
\def\bPsi{{\pmb\Psi}}
\def\hbk{\hat{\bf k}}
\def\bk{{\bf k}}
\def\hbe{\hat{\bf e}}
\def\bea{\begin{eqnarray}}
\def\ena{\end{eqnarray}}
\documentclass[aps,showpacs,prd,twocolumn,superscriptaddress]{revtex4}

\usepackage{amsmath}
\usepackage{epsfig,bm,epsf,float,amsbsy,ams, amssymb, cancel, lscape, rotating}
\usepackage{color}
\definecolor{bl}{cmyk}{1,1,0.3,0}
\usepackage[dvipdfm]{hyperref}

\begin{document}

%\preprint{}

\title{Constraints on the massive graviton dark matter from pulsar timing
and precision astrometry}


\author{Maxim Pshirkov}
\affiliation{Pushchino Radioastronomical Observatory, Lebedev
Physical Institute, Moscow, Russia} \email[]{pshirkov@prao.ru}
\author{Artem Tuntsov}
\affiliation{Sternberg Astronomical Institute, Moscow, Russia}
\author{Konstantin A. Postnov}
\affiliation{Sternberg Astronomical Institute, Moscow, Russia}

%\thanks{}
%\affiliation{}


\date{\today}

\begin{abstract}
The effect of a narrow-band isotropic stochastic GW background on
pulsar timing and astrometric measurements is studied. Such a
background appears in some theories of gravity. We show that the
existing millisecond pulsar timing accuracy ($\sim 0.2 \,\rm{\mu
s}$) strongly constrains possible observational consequences of
theory of massive gravity with spontaneous Lorentz braking
\cite{dtt:2005}, essentially ruling out significant contribution
of massive gravitons to the local dark halo density. The
present-day accuracy of astrometrical measurements ($\sim 100
\,\rm{\mu as}$) sets less stringent constraints on this theory.

\end{abstract}

\pacs{04.30.-w, 04.80.-y, 95.35.+d, 97.60.Gb,98.80.-k}
%\keywords{}
\preprint{}
\maketitle

\section{Introduction}

Spectacular progress in observational cosmology, especially in
measurements of CMB radiation, has challenged our
understanding of the Universe. The standard cosmological
$\Lambda$CDM model based on GR is
confirmed by observations with high accuracy \cite{wmap5, wmap5_1}. This model
requires the present Universe to be dominated by dark matter and
dark energy of unknown nature, so the modification of gravity at
large distances could provide alternative description of the
Universe. There are several theories with infrared modification of
gravity based on quite different grounds (e.g.
\cite{Milgrom:1983pn,Bekenstein:2004ne,Gregory:2000jc,Dvali:2000hr,Carroll:2003wy,Kogan:2000vb,Damour:2002ws,Arkani-Hamed:2003uy,Rubakov:2004eb}).
Among these possibilities, recently developed theories of massive
gravity with violated Lorentz invariance
\cite{Arkani-Hamed:2003uy,Rubakov:2004eb,Dubovsky:2004}
appear to be theoretically
attractive and have interesting phenomenology
(see the recent review \cite{Rubakov:2008}). In particular, in
theory of massive gravity \cite{Dubovsky:2004}, the Lorentz
invariance is spontaneously broken by the condensates of scalar
fields, which allows to avoid problems of strong coupling and
ghosts that are unavoidable in Lorentz-invariant theories with
massive graviton.

Dubovsky \cite{Dubovsky:2004} constructed a theory where
gravitational waves (GWs) are massive while linearized equations
for scalar and vector metric perturbations, as well as spatially
flat cosmological solutions, are the same as in GR. In this theory
an extra dark-energy term appears in the Friedmann equations
suggesting an unusual explanation to the observed accelerated
expansion. In addition, massive gravitons could be produced in the
early Universe copiously enough  to explain, in principle, all of
the cold dark matter  \cite{dtt:2005}. A distinctive feature of
GWs produced by cold massive gravitons is a very narrow frequency
range of the signal ($\Delta \nu/\nu \sim 10^{-6}$) as determined
by virial motions of cold gravitons in the galactic halo. The
central frequency itself is model-dependent, but GW emission from
known relativistic binary systems place an upper limit on the
frequency $\nu\le 3\times 10^{-5}\,\rm{Hz}$. At lower frequencies,
the amplitude of the GW background could be of order
$h\sim10^{-10}\left(\frac{3\times 10^{-5}\,\rm{Hz}}{\nu}\right)$
\cite{dtt:2005} assuming the density of gravitons matches the
conservative estimate of the  dark matter local density
$\rho_{DM}=0.3 \,\rm{GeV~cm^{-3}} $\cite{bgz:1992} . Clustering of
GWs on $\sim\mathrm{kpc}$ scales constrains de Broigle length of
massive graviton accordingly thereby placing a lower limit of
$\sim10^{-8}\,\mathrm{Hz}$. This leaves out a region
$\sim10^{-8}\,\mathrm{Hz}<\nu<\sim3\times10^{-5}\,\mathrm{Hz}$ for
the allowed frequency of GW associated with massive gravitons. The
amount of GW signal in the frequency range  $\sim10^{-5}-10^{-6}
\,\rm{Hz}$ is  further constrained by the tracking data for the
Cassini spacecraft \cite{armstrong:2003} (see Fig. \ref{figure}).

The aim of this  note is to show that the amount of (almost)
monochromatic GW in the entire allowed region can be strongly
constrained from the existing pulsar timing data and astrometric
measurements, essentially ruling out any significant contribution
due to massive gravitons to the  density of galactic dark matter.
The propagation of electromagnetic waves from a remote
astronomical source in the presence of a GW background causes an
excessive noise in pulsar timing \cite{sazhin:1978,
detweiler:1979} and alters stochastically the apparent position of
the source \cite{kaiser:1997,kopeikin:1999}. So, high-precision
pulsar timing and astrometry of distant sources (for example,
quasars) can be used to constrain the amplitude of the possible GW
background.

\section{Constraints from pulsar timing}
\label{sec:timing} Pulsar timing  was suggested in the late 1970s
\cite{sazhin:1978, detweiler:1979} as a tool  to detect or
constrain the local GW background. A GW travelling through the
Solar system affects the observed frequency of a pulsar resulting
in anomalous residuals in the time of arrival (ToA) of pulses
\cite{lorimer:2005}. Because of  unrivalled rotational stability,
timing of millisecond pulsars is particularly well suited for
detecting GWs  \cite{manchester:2007}. The conventional technique
of the stochastic background measurements using pulsar timing
\cite{jenet:2005} assumes correlating ToA residuals of several
pulsars. Using this method has yielded upper limits on the
low-frequency broad-band stochastic GW backgrounds
\cite{jenet:2006}.

A narrow-band GW background  produces an excessive noise in pulsar
timing at the corresponding frequency. The rms of timing residuals
of even a single pulsar can put an upper limit on the  GW
background amplitude in the frequency range between  the inverse
of the pulsar timing data time span $T$ (typically several years)
and inverse time of the pulsar signal accumulation($\sim$ hours).
The Parkes Pulsar Timing Array (PPTA) includes several pulsars
with current rms residuals $r\le 0.2\,\mathrm{\mu s}$
($0.12\,\mathrm{\mu s}$, $0.19\,\mathrm{\mu s}$ and
$0.17\,\mathrm{\mu s}$ for J0437-4715, J1713+0747¾ and J1939+2134,
respectively) \cite{manchester:2007}.

In the weak field limit of Dubovsky et al. theory, the equations
of motion are the same as in Einstein's GR and we can therefore
employ the results of GR calculations on the expected effect of a
local GW on the observed frequency of a pulsar. For the GW power
spectrum per logarithmic interval (as defined by Eq.~(18)
of~\cite{deepak1:2008}) we restrict ourselves to a $\delta$-like
function at some $k=2\pi\nu/c$: \bea \label{narrowspectrum}
P_h(k')= \left\{
\begin{tabular}{l}
$P_0,k<k'<k+\delta k$\\
0,~otherwise \\
\end{tabular} \right.
\ena
For this power spectrum, the mass density of GWs is~\cite{deepak1:2008}
\begin{equation}
\rho_\mathrm{GW}=(16\pi G)^{-1}c^2k P_0\delta k \label{P02rho}
\end{equation}
providing necessary connection to the total power $P_0\delta k$.

\begin{figure}
\center
\includegraphics[width=86mm]{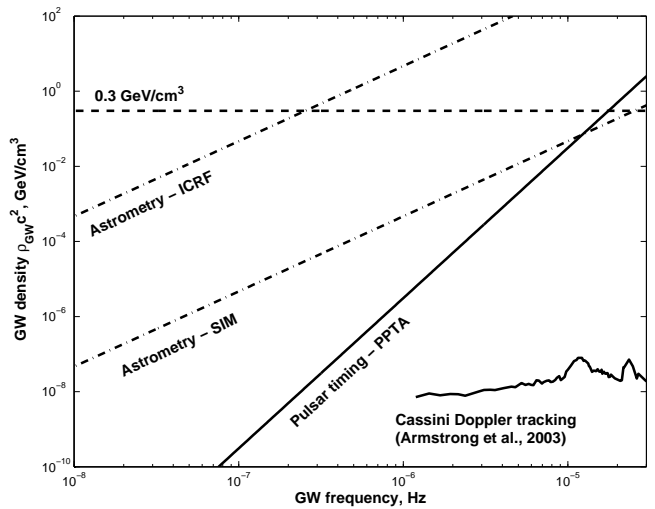}
\caption{Astrometric (dot-dashed) and pulsar timing (solid)
constraints on the overall energy density of a stationary
isotropic background of monochromatic GWs as a function of the
frequency $\nu$ in the range
$\sim10^{-8}\,\mathrm{Hz}<\nu<\sim3\times10^{-5}\,\mathrm{Hz}$
allowed by binary pulsar GW emission and DM clustering constraints
(see Introduction). The thick dashed line corresponds to the local
DM energy density of $0.3\,\mathrm{GeV}\mathrm{cm}^{-3}$. The
constraint in the lower right corner of the graph is set by the
Doppler tracking of the {\it Cassini} spacecraft
\cite{armstrong:2003}. } \label{figure}
\end{figure}

The observed pulsar ToA rms variation $r$ gives an upper limit on
ToA dispersion due to GWs and therefore translates into the
following upper limit on $P_0\delta k$ (see
Appendix~\ref{sec:appendix:timing}):
\begin{equation}
\label{power_spectrum_limits} P_0\delta k\leq3r^2k^3c^2,
\end{equation}
and therefore
\begin{eqnarray}
\label{gwdensity_limits12} \rho_{GW}c^2\leq(16\pi G)^{-1}3r^2k^4c^4=3G^{-1}\pi^3r^2\nu^4 \hspace{1cm}\\
\approx 2.5\mathrm{GeV}\cdot\mathrm{cm}^{-3}.
\left(\frac{\nu}{3\times10^{-5}\,\mathrm{Hz
}}\right)^4\left(\frac{r}{0.2\,\mathrm{\mu s}}\right)^2\nonumber
\end{eqnarray}
The upper limit corresponding to $r=0.2\,\mathrm{\mu s}$ is
plotted in Figure~\ref{figure} as a function of $\nu$ and is
clearly lower than what is  needed for massive gravitons to be the
dominant component of the local dark matter at regions
unconstrained by the Cassini data.

\section{Astrometric constraints}
\label{sec:astrometry} The astrometric effect is different for the
light propagating across a region of space with enhanced density
of massive gravitons, e.g. a dark halo of a galaxy or galaxy
cluster (the en-route effect), and for stochastic change of the
position of the observer immersed in the massive graviton halo
(the local effect). The former smears out the visible size of a
distant source, while the latter changes stochastically the
angular separation between different sources on the sky.

In astrophysically relevant cases the en-route effect is too small
to be detected at the present level of accuracy of astrometric
measurements. A very generous upper limit on the stochastically
fluctuating change in the observed position of a distant source
may be estimated as $\sigma_\Psi\sim
h_\mathrm{max}\Psi_\mathrm{L}$, where $\Psi_\mathrm{L}$ is the
angular size of the halo on the line of sight and $h_\mathrm{max}$
is the maximum amplitude of GWs comprising the halo (which, due to
the Gaussian nature of these fluctuations, is essentially the same
as the rms amplitude $h_c=\sqrt{P_h}$ that can be estimated from the dark matter
density).

Contrary to naive expectations, in GR the light ray deflection
does not execute a random walk and does not show the
$\sim\sqrt{N}$ growth of the deflection. Instead, for traceless
tensor perturbations travelling with the speed of light, only the
gravitational wave field at emission and detection points matter
\cite{Braginskyetal1990, kaiser:1997, Linder1986}. The relative
change $\Delta\Psi/\Psi$ in the angular separation between two
sources due to the local effect is also of order of the GW
background amplitude $h_c$. At $h_c\sim 10^{-11} - 10^{-10}$, as
Dubovsky~et~al. model suggests \cite{dtt:2005}, this would yield
$\sim\mathrm{\mu as}$ jitter in the angular separation for a
couple of sources across the sky. Such jitter can be discovered in
the future astrometric space experiments like SIM
\cite{Shao:1998}.

Present-day astrometric accuracy $\sigma_\Psi$ can be estimated as that
of the radio VLBI-based ICRF (International
Cosmic Reference Frame) \cite{Ma:1998, Fey:2004}, which involves
more than 200 reference radio sources determining the  celestial
coordinate frame. The ICRF sources are observed for many years,
and the accuracy of determination of source coordinates on the sky
relative to this frame may be used as a measure of the angular
separation stability. The best present-day accuracy of
$100\,\mathrm{\mu as}$ (at $1\,\sigma$ level) \cite{Zharov:2008}
means $\Delta\Psi\leq 5\times 10^{-10}$.

For the adopted power spectrum~(\ref{narrowspectrum}), this accuracy
translates into the following upper limit on the GW mass density (see Appendix~\ref{sec:appendix:astrometry}):
\begin{eqnarray}
\label{rhogwlimitastrometry} \rho_\mathrm{GW}\leq\frac{3\pi\nu^2\sigma_\Psi^2}{4G} \hspace{5cm}\\
\approx 4.2\times10^3\,\mathrm{GeV}\cdot\mathrm{cm}^{-3}
\left(\frac{\nu}{3\times10^{-5}\,\mathrm{Hz
}}\right)^2\left(\frac{\sigma_\Psi}{100\,\mathrm{\mu
as}}\right)^2. \nonumber
\end{eqnarray}
Other limits for this value as a function of frequency $\nu$ are
summarized on Fig.~\ref{figure}.

\section{Conclusions}
\label{sec:conclusions}

We have shown that the existing data on the millisecond pulsar
timing stability set a tight upper limit on the narrow-band GW
background amplitude at frequencies $\nu\leq10^{-5}\mathrm{Hz}$.
This limit can be used to severely  bound the amount of massive
cold gravitons which can potentially produce a strong narrow-band
GW background \cite{dtt:2005}.

The present-day astrometric constraints are less restrictive than
the timing ones at considered frequencies.  However, both are
still far above the tightest constraint set by the Doppler
tracking of Solar system spacecrafts  in the frequency range
$\nu>10^{-6}\,\mathrm{Hz}$ \cite{armstrong:2003}.

\begin{acknowledgments}
The authors acknowledge P.G. Tinyakov for useful discussion and
the anonymous referee for valuable comments. The work of M. P. is
supported by RFBR Grants No. 06-02-16816-a and No. 07-02-01034-a.
K.P. acknowledges partial support by the DAAD grant A/07/09400 and
grant RFBR 07-02-00961.
\end{acknowledgments}

%%%%%%%%%%%%%%%%%%%%%%%%%%%%%%%%%%%%%%%%%%%%%%%%%%%%%%%%%%%%%%%%%%%%
%\newpage
%%%%%%%%%%%%%%%%%%%%%%%%%%%%%%%%%%%%%%%%%%%%%%%%%%%%%%%%%%%%%%%%%%%%
\appendix

\section{Pulsar timing in narrow-band GW background}
\label{sec:appendix:timing}To calculate the effect of a
narrow-band GW background on pulsar timing one can use the
formalism presented in Baskaran et al. \cite{deepak1:2008}
starting with their expression~(9), which describes the effect of
a GW with wavevector $\bf{\tilde{k}'}$ and polarization tensor
$p_{ij}$ on the observed rotational frequency $f$ of a pulsar at
position $\hbe$ on the sky. Frequency fluctuation due to modes
with all $s$ and $\bk^\prime$  is
\begin{equation}
\Delta f(t)/f= \int\md\bk^\prime\, \left[h_s(\bk^\prime, t) g^s(p_{ij}, \hbk^\prime, \hbe) + \mathrm{c.c.}\right], \label{apptiming:start}
\end{equation}
where $g(p_{ij}, \hbe, \hbk^\prime)$ is given by~(9)
of~\cite{deepak1:2008} (setting the parameter $\epsilon\to0$) and
depends on the relative orientation of $\hbe,\hbk^\prime$ and
choice of the polarization gauge; in the above formula, summation
of circulary polarized modes~$s=1,2$ is implied and `c.c.' stands
for the complex conjugate.

The time of arrival residual $R(t)$ is the integral of relative fluctuation $\Delta f(t)/f$ w.r.t. time:
\bea R(t) = \int\limits_{0}^t\md t\Delta f(t)/f
\label{apptiming:residual1} \ena
and its autocorrelation function is
\bea
R_2(\tau)\equiv\lim\limits_{T\to\infty}T^{-1}\int\limits_0^T\mathrm{d}t\,R(t)R(t+\tau).
\ena
In particular, plugging the above expressions into each other and using
statistical properties of $h_s(\bk, t)$ as set by Eq.~(18) of~\cite{deepak1:2008}
one obtains for the ToA residual rms
\bea
r^2=R_2(0)=(3c^2)^{-1}\int\md k^\prime\,P_h(k^\prime)k^{\prime -3}
\ena
Applied to the power spectrum~(\ref{narrowspectrum}), this imediately
gives
\bea
P_0\delta k=3r^2k^3c^2
\ena
and corresponding limit~(\ref{gwdensity_limits12}) on the GW mass density.

\section{Astrometric fluctuations in narrow-band GW background}
\label{sec:appendix:astrometry}
To calculate the fluctuation $\Delta{\bf n}$ in the position of a source $\hbn$ we write
the astrometric effect due to a single GW mode in the formalism of~\cite{deepak1:2008}
similarly to Appendix~\ref{sec:appendix:timing}:
%writing the fluctuation $\Delta{\bf n}$ in the observed position of a source $\hbn$ as
\begin{equation}
\Delta{\bf n}(t)= {\bf f}^s(\hbk, \hbn) h_s(\bk, t) + \mathrm{c.c.}, \label{appastrometry:start}
\end{equation}
where conventions following~(\ref{apptiming:start}) are respected and factors
${\bf f}^s(\hat{\bf{k}}, \hat{\bf{n}})$ can be found by appropriately rotating
the results of Pyne et al. \cite{PyneGwinnetal1996}:

\begin{equation}
{\bf f}^{1,2}(\hat{\bf{k}}, \hat{\bf{n}})= \frac12\left\{\left[\hbn,\left[\hbk,\hbn\right]\right]\pm i\left[\hbn,\hbk\right]\right\}.
\end{equation}
As $\Delta\bn\perp\hbn$, the former has only two independent components.
For a pair of sources $\hbn, \hbn^\prime$ it is natural to choose these components to be along the great circle connecting the pair and perpendicular to it. Using
the above formulae and the statistical properties of $h_s(\bk, t)$ as introduced in Eq.~(18) of~\cite{deepak1:2008} one obtains for the correlation matrix
\begin{eqnarray}
\left(\begin{matrix}\Delta n^\parallel(\hbn, t)\Delta n^\parallel(\hbn^\prime, t^\prime) & \Delta n^\parallel(\hbn, t)\Delta n^\perp(\hbn^\prime, t^\prime) \cr \Delta n^\perp(\hbn, t)\Delta n^\parallel(\hbn^\prime, t^\prime) & \Delta n^\perp(\hbn, t)\Delta n^\perp(\hbn^\prime, t^\prime)\end{matrix}\right)=  \hspace{0.5cm} \label{DeltanMatrix} \\
= \frac16\left(\begin{matrix}1 & 0\cr 0& 1\end{matrix}\right)\cos^2\frac\Psi2 P_c(\tau) + \frac16\left(\begin{matrix}1 & 0\cr 0& -1\end{matrix}\right)\sin^2\frac\Psi2 P_c(2T)\nonumber \hspace{0cm}%\\
%= 2H_0^2\left(\begin{matrix}1 & 0\cr 0& 1\end{matrix}\right)\cos^2\frac\Psi2\sum\limits_{\omega\ge0}\Omega_\mathrm{GW}(\omega)\frac{\cos{\omega(t-t^\prime)}}{\omega^2}, \nonumber
\end{eqnarray}
where $\Psi=\angle(\hbn, \hbn^\prime)$, $\tau=t^\prime-t$, $2T=t^\prime+t$ and $P_c(t)$ is the cosine Fourier transform of the GW power spectrum $P_h$ (as defined by Eq.~(18) of~\cite{deepak1:2008}):
\begin{equation}
P_c(t)\equiv\int\md k\, k^{-1}P_h(k)\cos{ckt};
\end{equation}
for a $\delta$-like power spectrum~(\ref{narrowspectrum}), $P_c(t)=k^{-1}P_0\delta k\cos{ckt}$.

The autocorrelation function of the principal observable, the
fluctuation $\Delta\Psi$ is
\begin{equation}
\left\langle\Delta\Psi\left(t\right)\Delta\Psi\left(t^\prime\right)\right\rangle=\frac13\sin^2\frac\Psi2\left[P_c(\tau)-P_c(2T)\right].
\end{equation}
Independent observations average away the second term in the square brackets above in accord with the stationarity of the problem. In particular,
the rms value ($\tau=0$) in this case is
\begin{equation}
\left\langle\Delta\Psi^2\right\rangle=\frac13\sin^2\frac\Psi2P_c(0)=\frac{P_0\delta k}{3k}\sin^2\frac\Psi2;
\end{equation}
hence the estimate~(\ref{rhogwlimitastrometry}).

The gravitational wave field is Gaussian and so is its linear transform $\Delta\Psi(t)$ (cf.~\ref{appastrometry:start}). Therefore~(\ref{DeltanMatrix}) readily gives the respective distribution functions. The results generalize naturally to multiple observables. For a set of $m$ angular distances $\Psi_i(t_i)=\Psi_i+\Delta\Psi_i(t_i)$
between source pairs $(\hbk_i, \hbn_i)$, the probability density $\varphi(\Delta\bPsi)$ of the observable vector
$$
\Delta\bPsi\equiv\left(\Delta\Psi_1(t_1), ..., \Delta\Psi_{m-1}(t_{m-1}), \Delta\Psi_m(t_m)\right)^\mathrm{T}
$$
is
\begin{equation}
\varphi(\Delta\bPsi)=\frac{\exp\left[-\frac{1}{2}\Delta\bPsi^\mathrm{T} \bsM^{-1}\left(\hbn_1, ..., \hbk_m; t_1, ..., t_m\right)\Delta\bPsi\right]}{\sqrt{(2\pi)^m\det\bsM}}\label{mdistancedistribution}
 \nonumber
\end{equation}
with elements of $\bsM$ given by ($\Phi_{ij}=\angle(\hbn_i+\hbk_i,\hbn_j+\hbk_j)$):
\begin{equation}
M_{ij}=\frac13\sin\frac{\Psi_i}2\sin\frac{\Psi_j}2\cos\Phi_{ij}P_c(t_i-t_j). \label{distancecorrelations}\nonumber
\end{equation}

\end{document}